\documentclass{aa}

\makeatletter
\renewcommand*\aa@pageof{, page \thepage{} of \pageref*{LastPage}}
\makeatother

\usepackage{graphicx}
\usepackage[colorlinks=true, linkcolor=blue, citecolor=blue, filecolor=blue,
  urlcolor=blue, pdfencoding=auto, psdextra]{hyperref}
\usepackage[dvipsnames]{xcolor}
\usepackage{times}
\usepackage{hyperref}
\usepackage{orcidlink}

\begin{document}

\defcitealias{Habe-Ikeuchi-1980}{HI-80}

\title{Magnetohydrodynamic turbulence and the associated spatial \\
  diffusion tensor of cosmic rays in dynamical galactic halos}
\titlerunning{MHD turbulence in dynamical galactic halos}

\author{
  J.~Kleimann \thanks{Corresponding author: jk@tp4.rub.de}
  {}            \inst{1,2} \orcidlink{0000-0001-6122-9376} \and
  H.~Fichtner   \inst{1,2} \orcidlink{0000-0002-9151-5127} \and
  M.~Stein      \inst{3}   \orcidlink{0000-0001-8428-7085} \and
  R.-J.~Dettmar \inst{2,3} \orcidlink{0000-0001-8206-5956} \and
  D.J.~Bomans   \inst{2,3} \orcidlink{0000-0001-5126-5365} \and
  S.~Oughton    \inst{4}   \orcidlink{0000-0002-2814-7288}
}
\authorrunning{Kleimann et al.\ }

\institute{
  Ruhr-Universit\"at Bochum, Fakult\"at f\"ur Physik und Astronomie,
  Institut f\"ur Theoretische Physik IV, 44780 Bochum, Germany
  \and
  Ruhr Astroparticle and Plasma Physics Center (RAPP Center),
  44780 Bochum,Germany
  \and
  Ruhr-Universit\"at Bochum, Fakult\"at f\"ur Physik und Astronomie,
  Astronomisches Institut (AIRUB), 44780 Bochum, Germany            
  \and              
  Department of Mathematics, University of Waikato, 
  Hamilton 3240, New Zealand
}

\date{Received 23 January 2025 / Accepted 23 April 2025}

\abstract
    {A detailed understanding of cosmic-ray transport in galactic halos is essential for explaining various observations, such as the radio continuum measurements of synchrotron radiation from energetic electrons. Of central importance is the spatial diffusion tensor of cosmic rays, which can be computed in an ab~initio manner if the turbulence in the background medium is known.}
    {The study aimed to establish a suitable framework to compute the evolution of magnetohydrodynamic (MHD) turbulence and, hence, the diffusion tensor in the dynamical halos of galaxies.}
    {The Reynolds-averaged single-fluid MHD equations were solved numerically on a cylindrical grid, assuming axial symmetry and fixed boundary conditions on a central ellipsoid representing the galaxy. The physical properties of both large-scale MHD and small-scale turbulent quantities, including the coefficients of parallel and perpendicular diffusion, were extracted from the resulting steady state.}
    {Hydrodynamic validation revealed a persistent instability of the near-axis flow, which could be traced to the galaxy's gravitating mass. The results for the evolution of MHD turbulence in a galactic halo --~using parameters approximating those of \object{NGC~4631}~-- enabled an ab~initio computation of the spatial diffusion tensor of cosmic rays through the application of a state-of-the-art nonlinear theory.}
    {The simulation results reveal variation in turbulence quantities, (i.e., fluctuation energy, correlation scale, and cross helicity) in a dynamical halo. Furthermore, the corresponding diffusion tensor of cosmic rays exhibits significant variation throughout such a halo.}
    

\maketitle

\section{Introduction}

The physics of dynamical galactic halos, a term introduced almost 50 years ago \citep{Jokipii-1976, Cesarsky-1980, Lerche-Schlickeiser-1981}, has recently returned into the focus of contemporary research \citep[e.g.,][and references therein]{Bustard-etal-2016, Rupke-2018, Zhang-2018, Krause-2019, Recchia-2020, Recchia-etal-2021}.
New or upgraded radio telescopes such as the JVLA \citep{Perley-etal-2011} or LOFAR \citep{vanHaarlem-etal-2013} are collecting extensive new data, the interpretation of which is fostered through the development and application of advanced hydrodynamic (HD) and magnetohydrodynamic (MHD) simulation codes
\citep[e.g.,][]{Jacob-etal-2018, Chan-etal-2019, Dashyan-Dubois-2020, Schneider-etal-2020, Butsky-etal-2022, Tsung-etal-2023}. 
On the one hand, dynamical halos play a key role in galaxy formation and evolution \citep[e.g.,][]{Ruszkowski-etal-2017, Wang-etal-2020, vandeVoort-etal-2021, Girichidis-etal-2024}, as their properties directly influence the galactic mass budget through feedback and accretion a topic central to the ``infall'' versus ``outflow'' debate \citep[see][and references therein]{Faucher-Giguere-Oh-2023}. On the other hand, starburst-driven \citep[e.g.,][]{Martin-Fernandez-etal-2016} and cosmic-ray-driven wind models \citep[e.g.,][]{Recchia-etal-2016, Farber-etal-2018, Dorfi-etal-2019}
have been used in studies that attempt to explain phenomena such as the Fermi bubbles \citep{Mertsch-Petrosian-2019}, galactic wind termination shocks \citep{Merten-etal-2018, Aerdker-etal-2024}, so-called galactopauses \citep{Shull-Moss-2020} or anisotropic diffusion \citep{Pakmor-etal-2016}, and drifts of cosmic rays \citep{AL-Zetoun-Achterberg-2021}.
While dynamical halos most often are considered to exist in the form of supersonic winds, they could potentially also occur as subsonic expansions, which, analogous to the stellar case, have been termed ``galactic breezes'' \citep{Fichtner-Vormbrock-1997}. Very recently, the notion of galactic breezes and their impact on cosmic-ray transport has been examined in the context of the Fermi bubbles \citep[see][]{Taylor-Giacinti-2017, Giacinti-Taylor-2018, Tourmente-etal-2023}.

A detailed understanding of cosmic-ray transport in dynamical halos is obviously essential for future progress. On the observational side, the radio continuum observations of synchrotron radiation from cosmic-ray electrons provide key constraints on their properties and other physical parameters, such as the magnetic field strength and the pressure of the interstellar medium in galactic halos \citep[see][]{Heesen-2021, Irwin-2024}.
On the theoretical side, despite these advances in the numerical modeling of dynamical halos, further improvement is needed. As opposed to (semi-)analytical models describing a magnetized Galactic wind \citep[e.g.,][]{Fichtner-etal-1991a} or the large-scale Galactic magnetic field \citep[e.g.,][]{Beck-etal-2012, Jansson-Farrar-2012, Ferriere-Terral-2014, Henriksen-etal-2018, Kleimann-etal-2019, Shukurov-etal-2019} simulations still lack state-of-the-art MHD modeling of turbulence. Neither dynamo models \citep[e.g.,][]{Shukurov-etal-2019, Steinwandel-etal-2022}, wind models \citep[see][and references therein]{Dorfi-Breitschwerdt-2012,Zhang-2018, Thomas-etal-2023}, nor models of the streaming instability in dynamical halos 
\citep{Dorfi-etal-2019, Lazarian-Xu-2022} incorporate turbulent velocity and magnetic fields in a self-consistent, 3D, time-dependent manner that would allow for the simultaneous and explicit computation of the characteristic amplitudes of velocity and magnetic fluctuations. However, knowledge of these fluctuations is a prerequisite for the application of modern, sophisticated ab~initio theories of cosmic-ray transport, developed as state-of-the-art models for the heliospheric modulation of cosmic rays \citep{Engelbrecht-Burger-2013, Wiengarten-etal-2016, Moloto-Engelbrecht-2020}.
These theories quantify the spatial diffusion tensor of cosmic rays in terms of magnetic fluctuation amplitudes and, thereby, pave the way for fully self-consistent simulations of galactic halos that incorporate the dynamics of the turbulent thermal plasma, magnetic fields, and nonthermal energetic particles. 

The present paper addresses the MHD modeling of turbulence by establishing a framework under which an ab~initio approach to cosmic-ray transport in galactic halos becomes feasible. After describing the model equations in Sect.~\ref{sec:model_equations} and their numerical implementation and validation in Sect.~\ref{sec:model_validation}, the model is applied to an example disk galaxy in Sect.~\ref{sec:applications}. The distributions of turbulence and thermal plasma in the halo are first computed self-consistently.
To follow, the diffusion tensor of cosmic rays is determined from the computed fluctuation amplitudes. In the final Sect.~\ref{sec:summary}, we summarize and discuss all findings and provide a brief outlook on future work.  

\section{Model equations}
\label{sec:model_equations}

The study is based on a Reynolds decomposition of the MHD equations into averaged and fluctuating components of both the bulk velocity $\vec{U}+\delta\vec{u}$ and the magnetic field $\vec{B} + \delta\vec{b}$ \citep[e.g.,][]{Wiengarten-etal-2015, Usmanov-etal-2016, Kleimann-etal-2023}. Incompressible fluctuations were considered, thereby disregarding any density perturbations.
This approximation was justified by the considerations presented in 
\citet{Zank-Matthaeus-1992, ZankMatt93, BhattacharjeeEA98, HunanaZank10}, 
who demonstrate that in a supersonic plasma with plasma beta near unity, MHD fluctuations can be assumed to be nearly incompressible -- as indeed observed in the solar wind \citep[e.g.,][]{Roberts-etal-1987, HowesEA12-slow, Adhikari-etal-2015}. We also note that while the bulk plasma wind flow is mildly supersonic, the rms fluctuations in the local comoving frame are not, lending further support to our neglect of density perturbations.

\subsection{Large-scale equations}

The above decomposition results in the following large-scale equations:
\begin{align}
  &\partial_t \rho + \nabla \cdot (\rho\, \vec{U}) = 0,  \label{eq:conti} \\[0.1cm]
  &\partial_t (\rho\, \vec{U}) + \nabla \cdot \left[\rho\, \vec{U U} 
    + \left(p + \frac{|\vec{B}|^2}{2\mu_0} + p_\mathrm{w}\right)\tens{1} \right. \nonumber \\
    & \hspace*{0.5cm} \left. - \left(1+\frac{\sigma_D\mu_0\rho \, Z^2}{2 B^2}\right)\vec{B}\vec{B} \right] = -\rho \, \vec{g}, \label{eq:mom}
  \end{align}
  \begin{align}
    &\partial_t e + \nabla \cdot
  \left[ \left(e + p + \frac{\|\vec{B}\|^2}{2\mu_0} \right) \vec{U}
    - \frac{(\vec{U} \cdot \vec{B}) \vec{B}}{\mu_0} - \frac{\rho H_\mathrm{c}}{2}\vec{V}_\mathrm{\!\!A} \right] \nonumber \\
  &\hspace*{0.5cm} = -\rho\,\vec{U}\cdot\vec{g} - \vec{U}\cdot\nabla p_\mathrm{w} - \frac{H_\mathrm{c}}{2}\vec{V}_\mathrm{\!\!A}\cdot\nabla\rho + \frac{\rho Z^3 f^+}{2\lambda}\nonumber \\
  &\hspace{0.8cm} +\ \vec{U}\cdot(\vec{B}\cdot\nabla)\left[\frac{\sigma_D \, \rho Z^2}{2B^2}\vec{B}\right] - \rho\vec{V}_\mathrm{\!\!A}\cdot\nabla(H_\mathrm{c}), \label{eq:ene}\\
  &\partial_t \vec{B} + \nabla \cdot (\vec{U}\vec{B}-\vec{B}\vec{U}) = \vec{0} \label{eq:indu} 
\end{align}
for the mean values of the plasma mass density $\rho$, the rest frame velocity $\vec{U}$, the total energy density given by
\begin{equation}
  e = \frac{\rho \, U^2}{2} + \frac{B^2}{2\mu_0} + \frac{p}{\gamma-1} , 
\end{equation}
and the magnetic field $\vec{B}$. The turbulence-related quantities
\begin{align}
  Z^2 &=
  \left<\delta\vec{u}^2\right> + \left<\delta\vec{b}^2/(\mu_0 \rho)\right>, \\
  \sigma_D \, Z^2 &=
  \left<\delta\vec{u}^2\right> - \left<\delta\vec{b}^2/(\mu_0 \rho)\right>, \\
  H_\mathrm{c} &=
  2 \, \left<\delta\vec{u}\cdot\delta\vec{b}/(\mu_0 \rho)\right>
\end{align}
denote the so-called ``wave energy density'' (equal to twice the turbulent energy density per unit mass), the ``residual energy'' (where the dimensionless scalar $\sigma_D$ is the normalized residual energy), and the cross helicity, respectively. Furthermore, $\lambda$ denotes the correlation length of the fluctuations, $p$ the thermal pressure, $p_\mathrm{w} = (\sigma_D+1)\rho Z^2/4$ the ``wave'' pressure, $\vec{V}_\mathrm{\!\!A}$ the large-scale Alfv\'en velocity, $\mu_0$ the permeability constant, and $\vec{g}$ the gravitational acceleration (see Sect.~\ref{sec:gravpot}).

\subsection{Small-scale equations}
\label{sec:smallscale}

The evolution of the turbulence quantities $Z^2$, $H_\mathrm{c}$, and $\lambda$ is governed by the following resulting small-scale equations:
\begin{align}
  &\partial_t Z^2 + \nabla\cdot(\vec{U}Z^2 + \vec{V}_\mathrm{\!\!A} H_\mathrm{c}) \nonumber \\
  &\hspace{0.3cm} = \frac{Z^2(1-\sigma_D)}{2}\nabla\cdot\vec{U} + 2(\vec{V}_\mathrm{\!\!A}\cdot\nabla) H_\mathrm{c} \nonumber \\
  &\hspace{0.7cm} + \frac{Z^2\sigma_D}{B^2}\vec{B}\cdot(\vec{B}\cdot\nabla)\vec{U}
  - \frac{\alpha Z^3f^+}{\lambda} + S, \label{eq:Z2}\\
  &\partial_t H_\mathrm{c} + \nabla\cdot(\vec{U} H_\mathrm{c} - \vec{V}_\mathrm{\!\!A} Z^2) \nonumber \\ 
  &\hspace{0.3cm} = \frac{H_\mathrm{c}}{2}\nabla\cdot\vec{U} + Z^2(\sigma_D-2)\,\nabla\cdot\vec{V}_\mathrm{\!\!A}
  - \frac{\alpha Z^3 f^-}{\lambda}, \label{eq:cross} \\
  &\partial_t(\rho\lambda) + \nabla\cdot(\vec{U}\rho\lambda) 
  = \rho\beta\left(Z f^+ - \frac{\lambda}{\alpha Z^2}S\right), \label{eq:lam}
\end{align}
with the K\'arm\'an--Taylor constants $\alpha$ and $\beta$ \citep[e.g.,][]{Wiengarten-etal-2015, Usmanov-etal-2016}, also referred to as K\'arm\'an--Howarth constants \citep[e.g.,][]{Pei-etal-2010, Kleimann-etal-2023}.
The function $S$ in Eqs.~(\ref{eq:Z2}) and (\ref{eq:lam}) can be used to describe sources of turbulence in a halo. The auxiliary functions $f^\pm$ in both sets of equations are defined by
\mbox{$f^\pm = \sqrt{1-\sigma_\mathrm{c}^2}\, \left(\!\sqrt{1+\sigma_\mathrm{c}}\pm\!\sqrt{1-\sigma_\mathrm{c}}\right)/2$} with the normalized cross helicity given as $\sigma_\mathrm{c} = H_\mathrm{c}/Z^2$. As opposed to the latter, the normalized energy difference $\sigma_D$ was approximated as a global constant in each simulation presented in this work (but see Sect.~\ref{sec:sigma_D}).

By definition, both $\sigma_\mathrm{c}$ and $\sigma_D$ are not only individually restricted to the interval $[-1,1]$, but also need to fulfill the condition \citep[e.g.,][]{Perri_Balogh:2010}
\begin{equation}
  \label{eq:sigma_circle}
  \sigma_\mathrm{c}^2 + \sigma_D^2 \le 1
\end{equation}
that follows directly from their respective definitions.
However, since neither constraint is ensured by Eqs.~(\ref{eq:Z2})--(\ref{eq:lam}), simulations often require $|\sigma_\mathrm{c}| \le \sqrt{1-\sigma_D^2}$ to be explicitly enforced at runtime.

Within contemporary theory, turbulence should be considered anisotropic relative to a significant magnetic background (or guide) field. In particular, a useful (over)simplification is to assume, as in
\citet{Matthaeus-etal-1990}, that the magnetic fluctuations consist of two components (i.e., \mbox{$\delta\vec{b} = \delta\vec{b}_\mathrm{sl} + \delta\vec{b}_\mathrm{2d}$}): a slab component $\delta\vec{b}_\mathrm{sl}$ (with wave vectors along the magnetic field) and a quasi-two-dimensional component $\delta\vec{b}_\mathrm{2d}$ (with wave vectors perpendicular to\ the field). These components determine the diffusive transport of cosmic rays along and perpendicular to the magnetic field, respectively (see Sect.~\ref{sec:diffuse} below).

Equations~(\ref{eq:Z2})--(\ref{eq:lam}) represent a so-called one-component model for turbulence evolution, which is constructed to describe only the quasi-two-dimensional fluctuations \citep[e.g.,][]{Smith-etal-2001, Minnie-etal-2005, Breech-etal-2008, Usmanov-etal-2016, Kleimann-etal-2023}, i.e., $\delta\vec{b} = \delta\vec{b}_\mathrm{2d}$. The energy density $Z^2$ can then be translated into $\delta b_\mathrm{2d}^2$ via \citep{Roberts-etal-1987, Wiengarten-etal-2016}
\begin{equation}
  \label{eq:delta_b2}
  \delta{b}_\mathrm{2d}^2 \equiv \left< \delta \vec{b}^2 \right>
  = \frac{1-\sigma_D}{2}\mu_0 \rho \, Z^2 .
\end{equation}
To obtain values for $\delta b^2_\text{sl}$, we followed a standard procedure
\citep[see][]{Minnie-etal-2005, Pei-etal-2010, Chhiber-etal-2017}
where the ratio $\delta{b}_\mathrm{sl}^2 / \delta{b}_\mathrm{2d}^2$ is assumed constant.
We note that more sophisticated turbulence models have been used in solar wind applications
\citep[e.g.,][]{Oughton-etal-2011, Engelbrecht-Burger-2013, Wiengarten-etal-2016, Adhikari-etal-2017, Moloto-Engelbrecht-2020}, and similar models could be employed for galactic halos in future studies.

\subsection{The gravitational potential}
\label{sec:gravpot}

The gravitational acceleration $\vec{g}$ in the above large-scale equations can be derived from the total gravitational potential $\Phi$ of a galaxy via $\vec{g} = -\nabla\Phi$. Contemporary modeling assumes $\Phi$ to consist of contributions from a bulge, a disk, and a spherical (dark-matter) halo \citep[see][and references therein]{Bajkova-Bobylev-2016}. The most commonly employed representation \citep[for recent examples, see][]{ Ramos-Martinez_EA:2018, Schneider-etal-2020} of the first two components is the axisymmetric one introduced by \citet{Miyamoto-Nagai-1975} using cylindrical coordinates $(R,z)$ in
\begin{equation}
  \label{eq:miyamoto_nagai}
  \Phi_j = - \frac{G M_j}{\sqrt{R^2 + \left(a_j + \sqrt{z^2 + b_j^2}\right)^2}}
\end{equation}
with $j=1$ for the bulge and $j=2$ for the disk component, where $z$ is the (possibly negative) height above the central plane of the disk and $R$ is the distance to the symmetry axis. $G$ is the gravitational constant, and $M_j$ are the respective masses of the bulge and the disk. The constants $a_j$ and $b_j$ are obtained from fits to a given galaxy.

The dark-matter halo potential is usually assumed to be radially symmetric.  
In the past, the potential attributed to \citet{Innanen-1973} was frequently used \citep[e.g.,][]{Habe-Ikeuchi-1980, Breitschwerdt-etal-1991, Zirakashvili-etal-1996}, as
\begin{equation}
  \label{eq:innanen}
  \Phi_\mathrm{halo} = -\frac{G M_\mathrm{halo}}{r_\mathrm{b}}\left(\ln (1+r/r_\mathrm{b}) + \frac{1}{1+r/r_\mathrm{b}}\right),
\end{equation}
where $r=\sqrt{R^2+z^2}$ is the galactocentric distance, $r_\mathrm{b}$ is a reference radius, and $M_\mathrm{halo}$ is the mass of the halo. 
In recent years, a potential corresponding to a Navarro-Frenk-White \citep{Navarro-etal-1997} halo mass distribution given by
\begin{equation}
  \label{eq:nfw}
  \Phi_\mathrm{halo} = -\frac{G M_\mathrm{halo}}{r}\,\ln\left(1+\frac{r}{r_{200}/c_\mathrm{h}}\right)    
\end{equation}
has gained widespread use. Here, $M_\mathrm{halo}$ and the ``concentration parameter'' $c_\mathrm{h}$ are chosen such that the mass contained inside a sphere of radius $r_{200}$ is given by $M_{200} = M_\mathrm{halo} \left[ \ln(1+c_\mathrm{h})-c_\mathrm{h}/(1+c_\mathrm{h}) \right]$. Owing to Newton's theorem, the fact that the total mass is infinite remains inconsequential for typical applications. Additional models are discussed in, for example, \citet{Granados-etal-2021}.

\section{Model validation}
\label{sec:model_validation}

\subsection{Numerical implementation}
\label{sec:numerics}

We used the finite-volume MHD and multifluid code Cronos \citep{Kissmann-etal-2018} 
to solve the extended MHD equations~(\ref{eq:conti})--(\ref{eq:lam}), amended by an additional equation for the entropy density $p/\rho^{\gamma}$ to ensure pressure positivity \citep{Balsara_Spicer:1999}. The number of cells in which this so-called ``entropy fix'' is applied --~and where the energy equation~(\ref{eq:ene}) is thus temporarily disregarded~-- typically peaks at a few percent of the total volume in the early phase of a simulation, and consistently decreases to zero long before the final steady state is attained. We employed a two-dimensional cylindrical grid
\mbox{$(R,z) \in [0,24]$\,kpc$\times [0,14]$\,kpc} with a uniform grid spacing of $\Delta R = \Delta z=40$\,pc. Extending the grid into the azimuthal \mbox{($\varphi$-)direction}, while technically straightforward, would require considerably more computational resources. This step is not warranted for the present study, but could be useful in a follow-up paper investigating nonaxisymmetric effects, such as spiral arms or triaxial halos.

The galaxy itself is represented by an oblate ellipsoid with semi-axes $R_\mathrm{c}$ and $z_\mathrm{c}$, on which all quantities except the magnetic field are prescribed as time-independent Dirichlet boundary conditions.
In particular, the poloidal velocity was kept at $U_R=U_z=0$ within the galactic ellipsoid.
The reason for this latter choice is that any nonzero poloidal velocity would, via the induction Equation~(\ref{eq:indu}), cause the poloidal magnetic field inside the ellipsoid to depart from its desired boundary condition. Restoring the internal magnetic field to its initial value would then inevitably cause an undesirable layer of nonzero $\nabla\cdot\vec{B}$ at the ellipsoid's surface. However, by enforcing a purely toroidal velocity within the central ellipsoid, the unconditional validity of the induction equation then yields
\begin{equation}
  \label{eq:pol_induct}
  \partial_t \vec{B}_\mathrm{pol} = \nabla \times (U_\varphi \, \vec{e}_\varphi
  \times \vec{B}_\mathrm{pol}) = \nabla \times \vec{0} = \vec{0} ,
\end{equation}
thereby keeping the in-disk poloidal magnetic field $\vec{B}_\mathrm{pol}$ equal to its initial value at all later times.
This also ensures that the \mbox{$\nabla \cdot \vec{B}=0$} constraint is maintained to machine accuracy throughout the simulation, simply by virtue of the induction equation~(\ref{eq:indu}) and the constrained-transport scheme employed in Cronos \citep[see, e.g.][]{Evans_Hawlay:1988, Balsara_Spicer:1999_staggered}, thereby eliminating the need to implement a dedicated divergence cleaning method.

The use of a prescribed internal boundary surface, which lacks an obvious physical counterpart in a real galaxy, could be criticized as artificial, particularly in the light of other works \citep[e.g.,][]{Fielding_EA:2017, Ramos-Martinez_EA:2018, Schneider-etal-2020} in which the galaxy is implemented simply through a gravitational well in which plasma orbits self-consistently due to its initial angular momentum and with optional large-scale outflows being driven through corresponding source terms. 
Although our method does entail some undesirable freedom and lack of self-consistency, we argue that it remains necessary for the following reason.
As shown by the simulations presented in Sects.~\ref{sec:model_validation} and~\ref{sec:applications}, the galactic winds obtained typically exhibit a radial component of significant magnitude. Without an internal region where $U_R=0$ is enforced, this outflow would likely also occur within the galaxy and transport the initial magnetic field outward. This would cause the galaxy, and thus its halo, to demagnetize quickly --~on the fluid crossing timescale~-- everywhere except possibly in the immediate vicinity of the rotational axis, where $U_R=0$ by symmetry. And unlike the mass and energy lost to the wind, the magnetic field would be very difficult to continuously replenish through respective source terms, at least if the divergence constraint is to be maintained. Indeed, none of the aforementioned studies includes both a magnetic field and a large-scale outflow.
(The inclusion of an actual small-scale dynamo, which would be the preferred way to physically resolve this issue \citep[e.g.,][]{Pfrommer_EA:2022}, is beyond the scope of our present work.)
For these reasons, we argue that the use of an internal boundary, however artificial on principal grounds, is indeed appropriate for the purpose at hand.

As initial conditions, we typically adopt globally constant values for $Z^2$, $\lambda$, $\sigma_\mathrm{c}=0$, and temperature $T$, while $\rho$ decreases $\propto r_\mathrm{e}^{-2}$ with ``elliptic'' distance, defined by
\begin{equation}
  \label{eq:r_ell}
  r_\mathrm{e}(R,z):= \sqrt{(R/R_\mathrm{c})^2+(z/z_\mathrm{c})^2}
\end{equation}
and thus measured from the interior boundary surface at $r_\mathrm{e}=1$. As is standard, the number density $n$ is linked to $\rho$ via $\rho=\mu\, m_\mathrm{p}\, n$, where $m_\mathrm{p}$ is the proton rest mass and the molecular weight is $\mu=0.62$, assuming solar metallicity. The azimuthal velocity is initialized as
\begin{equation}
  \label{eq:v_Kepler}
  U_\varphi|_{t=0} = \sqrt{R \ \partial_R (\Phi_1+\Phi_2+\Phi_\mathrm{halo})} ,
\end{equation}
that is, as the Keplerian velocity corresponding to the total gravitational potential.
The initial poloidal velocity chosen is purely radial with a sonic Mach number of
\begin{equation}
  \label{eq:v_init}
  \left. \sqrt{\frac{ U_R^2 + U_Z^2}{\gamma \, p/ \rho}} \right|_{t=0}
  = \left[ \mathrm{max} \left(0, 
    \frac{R-R_\mathrm{c}}{R_\mathrm{max}-R_\mathrm{c}},
    \frac{z-z_\mathrm{c}}{z_\mathrm{max}-z_\mathrm{c}}
    \right) \right]^2 ,
\end{equation}
thus ensuring that the flow is supersonic at the outer boundaries ($R=R_\mathrm{max}$ and $z=z_\mathrm{max}$) and consistent with the inner boundary conditions. Zero poloidal velocity would be a possible alternative;
however, in some cases (notably in HI-ext, see Sect.~\ref{sec:hi_ext}) this was observed to result in the formation of a spurious vortex at larger $R$, which persisted for a long time before finally disappearing, thereby unnecessarily lengthening the simulation time spent to reach convergence. An illustration of such a vortex is available via the nonrotating test case depicted in Fig.~\ref{fig:quad_Uz}c, where it appears despite the initial condition~(\ref{eq:v_init}) adopted.
We deliberately refrained from using initial conditions based on hydrostatic equilibrium considerations \citep[e.g.,][]{Ramos-Martinez_EA:2018} because these would, by construction, not be conducive to the development of a large-scale wind.

For the magnetic field, several options are used, the simplest being a homogeneous vertical field, $\vec{B}|_{t=0}=B_0\, \vec{e}_z$. A slightly more sophisticated choice consists of hyperbolic field lines (see Appendix~\ref{app:hyper_B}), which have the convenient property of always being normal to the ellipsoid, regardless of the latter's aspect ratio $z_\mathrm{c}/R_\mathrm{c}$. 
Additional technical details on the boundary treatment are available in Appendix~\ref{app:boundary}. Table~\ref{tab:parameters} summarizes the physical parameters and initial conditions for all the simulations presented in this work.
Time integration is typically halted after 200\,Ma to 300\,Ma, at which point the system has either reached a stationary state, or it has become apparent that the system will not become stationary at all. An illustrative example of such behavior is available in Sect.~\ref{sec:habe_ikeuchi}. 

\subsection{Comparison with the Habe \& Ikeuchi model}
\label{sec:habe_ikeuchi}

\begin{figure*}
  \centering
  \includegraphics[width=\linewidth]{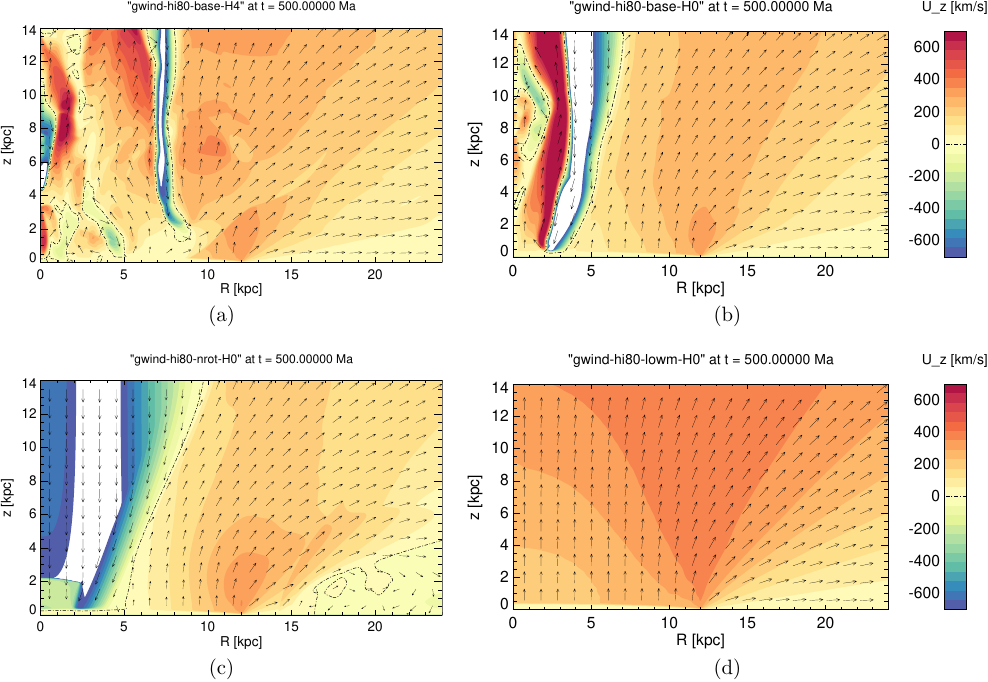}
  \caption{
    Snapshots at $t=500$\;Ma for four variations of our \citetalias{Habe-Ikeuchi-1980} benchmark simulations (``HI-base''), showing contours of $U_z$ with vector arrows of the poloidal velocity overplotted.
    Arrow length is proportional to $(U_R^2+U_Z^2)^{0.1}$ (here and elsewhere in this paper, unless otherwise stated), allowing the flow direction to be clearly discerned even in regions of low speeds.    
    (a) Standard case. (b) As in (a), but with the galactic disk extending to the $z$-axis, thereby ``closing'' the central 4\,kpc hole.
    (c) As in (b), but for a nonrotating disk.
    (d) As in (b), but with all gravitating masses ($M_{1,2}$, $M_\mathrm{halo}$) reduced to one tenth of their original values. White regions indicate values of $|U_z|$ in excess of 700\,km/s.
    The temporal evolution is available as an \textbf{ancillary file}.}
  \label{fig:quad_Uz}
\end{figure*}

The galactic wind simulations of \citet[][hereafter \citetalias{Habe-Ikeuchi-1980}]{Habe-Ikeuchi-1980} provide an ideal reference for our study of outflow from an axisymmetric disk galaxy, owing to their similar approach and the relative simplicity of these early models. These authors numerically solved the HD equations (equivalent to Eqs.~(\ref{eq:conti})--(\ref{eq:ene}) for \mbox{$\vec{B}=\vec{0}$} and $Z^2=H_\mathrm{c}=0$), amended by a ``cooling'' term (actually a negative source term in the energy equation) $-\rho^2 \Lambda(T)$ with $\Lambda(T)$ taken from \citet{Raymond_EA:1976}. This term was included for consistency but is ineffective for the low densities considered here. Their numerical setting is identical to that described in Sect.~\ref{sec:numerics}, except that the galaxy is represented not by an ellipsoid but by an annular disk, $R\in [4,12]$\,kpc in the \mbox{$(z=0)$} plane. We aimed to reproduce their ``wind-type'' model using the parameters listed in the first column (``HI-base'') of Table~\ref{tab:parameters}. For this setting, the authors report convergence into a stationary, supersonic wind on a timescale of about 200\,Ma.
\begin{table}
  \caption{Simulation parameters.}
  \label{tab:parameters}
  \centering
  \begin{tabular}{c@{\hspace*{2mm}}c|ccr@{\hspace*{1mm}}l}
    \hline\hline \rule{0mm}{3mm}
    Quantity & Unit & HI-base & HI-ext & N4631 & [Src.] \\
    \hline \rule{0mm}{3.5mm}
    $R_\mathrm{c}$ & kpc & 12.0 & 12.0 & 11.7 & [a]\\
    $z_\mathrm{c}$ & kpc & 0.25 &  1.0 & 1.0 & [b] \\
    \hline \rule{0mm}{3.5mm}
    $M_1$ & $10^{10}\, M_\sun$ & 2.050  & 0.205 & 0.0 & [c] \\
    $M_2$ & $10^{10}\, M_\sun$ & 25.47 & 2.547 & 1.585 & [c] \\
    $a_1$ & -- & 0.000 & 0.000 & \multicolumn{2}{c}{--} \\
    $a_2$ & -- & 7.258 & 7.258 & 3.0 & [c] \\
    $b_1$ & -- & 0.495 & 0.495 &  \multicolumn{2}{c}{--} \\
    $b_2$ & -- & 0.520 & 0.520 & 0.28 & [c]\\
    $M_\mathrm{halo}$ & $10^{10}\, M_\sun$ & 13.5 & 1.35 & 40.0 & [d] \\
    halo type & -- & Eq.~(\ref{eq:innanen}) & Eq.~(\ref{eq:innanen}) &
    \multicolumn{2}{c}{Eq.~(\ref{eq:nfw})} \\
    \hline \rule{0mm}{3.5mm}
    $n_\mathrm{disk}$ & $10^3$ m$^{-3}$ & 1.0 & 1.0 & 2.8 & [c] \\
    $T_\mathrm{disk}$ & MK & 5.0 & 5.0 & 5.6 & [e] \\
    $B_\mathrm{disk}$ & nT & -- & 0.1 & 1.0  & [f] \\
    $L_B$             &kpc & -- & $\infty$ & 8.0 & [g] \\
    field lines       & -- & homog. & homog. & \multicolumn{2}{c}{hyperbolic} \\
    cooling           & -- &  yes   &  yes   & \multicolumn{2}{c}{no} \\
    \hline \rule{0mm}{3.5mm}
    $Z^2_\mathrm{disk}$ & (km/s)$^2$ & -- & 5500 & \multicolumn{2}{c}{5500} \\
    $\lambda_\mathrm{disk}$ & kpc & --  & 0.1  & 0.1 & [h]\\
    $\gamma$                & -- & 1.67 & 1.67 & \multicolumn{2}{c}{1.40} \\ 
    \hline
  \end{tabular}
  \tablefoot{Simulation parameters of the boundary layout, gravitational potential~(\ref{eq:miyamoto_nagai}), and galactic boundary conditions for large-scale and small-scale quantities adopted in this study, or more specifically the \citetalias{Habe-Ikeuchi-1980} verification (``HI-base,'' Sect.~\ref{sec:habe_ikeuchi}), its low-mass MHD-turbulent extension (``HI-ext,'' Sect.~\ref{sec:hi_ext}), and the more realistic analog to \object{NGC~4631} (``N4631,'' Sect.~\ref{sec:refgal}).
    For HI-base, $z_\mathrm{c}$ was chosen to match the grid cell size used by \citetalias{Habe-Ikeuchi-1980}. Values without sources are justified in the main text. }
  \tablebib{
    [a]~\citet{Wiegert_EA:2015}; [b]~\citet{HongBae_EA:2011};
    [c]~\citet{Wang_EA:1995}; [d]~\citet{Martinez-Delgado_EA:2015};
    [e]~\citet{Wang_EA:2023}; [f]~\citet{Beck-etal-2019};
    [g]~\citet{BasuRoy-2013}; [h]~\citet{Haverkorn-etal-2008},
    \citet{Houde-etal-2013}, \citet{Chamandy-Shukurov-2020},
    \citet{Seta-Federrath-2024}. } 
\end{table}
Surprisingly, our simulation did not settle into a steady state but instead developed a highly dynamic, large-scale Kelvin-Helmholtz-like instability near the axis. This consisted of strongly variable flows lacking any apparent
periodicity or ordered spatial structure, except for columns of strong vertical inflow that formed repeatedly near the $z$-axis on a timescale of roughly 600\,Ma and persisted for several 100\,Ma before gradually moving outward toward the $R_\mathrm{max}$ grid boundary. 
A consistent outflow was established only at larger radial distances, except for occasional disturbances caused by vortex-like features expelled from the central region (cf.~Fig.~\ref{fig:quad_Uz}a).

We conducted several tests to identify possible causes of this unexpected phenomenon. First, we performed a rerun with identical parameters on a fully 3D Cartesian grid to rule out numerical artifacts induced by the coordinate singularity at $R=0$. Here, the aforementioned behavior was reproduced within the expected accuracy, given finite cell sizes.
Next, the central hole in the annular disk, which appears to be of dubious physical relevance anyway, was closed by specifying inner boundary conditions over the full radial interval $[0,12]$\,kpc.
The instability persisted but was characterized primarily by more regular up- and downflows, rather than by a network of smaller vortices, as in the previous case (compare~Fig.~\ref{fig:quad_Uz}b). 
The tendency for vertical columns of inflowing material to form and migrate radially outward was again observed.

One might speculate that the attachment of these columns to the disk is an artifact of the disk's $\vec{U}=\vec{0}$ boundary condition, which is required to maintain a divergence-free $\vec{B}$ field (compare Sect.~\ref{sec:numerics}) and that any nonzero $U_z|_{z=0}$ would likely cause the inflow column to quickly move away from the disk. However, closer inspection of the dynamics, as revealed in an animation of the full run (available as an \textbf{ancillary file}), shows that the column persists even when temporarily disconnected from the disk. This suggests that it likely would also form if a nonzero velocity were imposed at the disk surface.

In an additional test run, we disabled Keplerian rotation~(\ref{eq:v_Kepler}) to investigate the relevance of centrifugal effects to the instability. The inflow column still formed but remained fixed to the $z$-axis, oscillating quasiperiodically in radial extent and exhibiting much higher speeds ($\sim1500$\,km/s) than before (compare~Fig.~\ref{fig:quad_Uz}c).
This indicates that the instability originates at the $z$-axis and is then transported to larger radii through centrifugal forces. 
We note, however, that its magnitude was likely amplified by the zero-gradient boundary conditions imposed at $z=z_\mathrm{max}$, which can sustain a spurious self-reinforcing inflow.

Lastly, we restored Keplerian rotation but reduced the masses of all the sources of gravity (disk, bulge, and halo) to 10\% of their original values. This caused the simulation to converge toward a steady supersonic wind after approximately 150\,Ma, with no indication of instability (compare~Fig.~\ref{fig:quad_Uz}d).
Our preliminary conclusions are therefore:
(i) the instability is a real effect; and 
(ii) it only arises in sufficiently massive galaxies, presumably because stronger gravity prevents upward-moving fluid elements from reaching escape velocity near the axis, while at larger distances from the axis, increasing centrifugal acceleration helps to push the fluid away from the disk.
Consequently, the flow configuration depicted in panel~b (d) of Fig.~\ref{fig:quad_Uz} can, in the framework of this relatively simple HD model, be considered representative of a high-mass (low-mass) galaxy, while panels~a and c are less realistic due to the annular hole in the disk (case~a) and the lack of rotation (case~c), respectively.
The question of why this behavior was not reported in \citetalias{Habe-Ikeuchi-1980} is tentatively addressed in Appendix~\ref{app:hi80-steady}.

The persistent irregular motion of small cloud-like blobs seen near the rotational axis is likely relevant for more massive galaxies such as the Milky Way. Indeed, \citet{DiTeodoro_EA:2018} observed extended clumpy structures composed of neutral hydrogen clouds above and below the Galactic center, while \citet{Marconcini_EA:2023} presented similar observations in three nearby Seyfert-II galaxies (\object{NGC 4945}, \object{NGC 7582}, and \object{ESO 97-G13}). Although both groups of authors interpreted their observations within the framework of a biconical radial flow of constant velocity, the data may also be consistent with less regular flow patterns --~including localized inflow regions~-- once the ad~hoc assumption of a constant and outward-directed flow is dropped.
Another piece of observational evidence for inward gas motions in the central regions of outflows from massive galaxies stems from Na~I interstellar absorption line profiles based on high-dispersion spectra
\citep[i.e.,][]{Rupke_EA:2002}.
In this study, two out of the eleven ultraluminous infrared galaxies (ULIRGs) show not only blueshifted absorption (the tell-tale signature of outflowing gas) but also redshifted velocity components consistent with the results of our simulations. One caveat is that these absorption lines could also originate from gas clouds remaining from the merger process that formed the ULIRG, but no further observational evidence supports this interpretation \citep{Rupke_EA:2002}. 
We intend to investigate the exact mechanism of this interesting phenomenon in the near future.

\subsection{Extension to turbulent MHD}
\label{sec:hi_ext}

Once we established that the low-mass version of \mbox{HI-base} allows the halo to settle into a stable outflow configuration, we proceeded to extend the model to full MHD, including turbulence.
As shown in the corresponding ``HI-ext'' column in Table~\ref{tab:parameters}, we retained all the nonzero parameters from the low-mass HI-base case, except for the disk thickness, which we increased by a factor of four. This facilitates a clearer characterization of the behavior of the turbulence quantities, and also allowed us to verify the magnetic coupling within the ellipsoid. Moreover, we used a horizontal magnetic field of uniform strength with a magnitude small enough to remain inside the high-beta regime throughout the simulation domain. 
As a result, the HD quantities were not expected to differ significantly from those of the purely hydrodynamic HI-base run.

We kept this first application of our model for the evolution of MHD turbulence in a galactic halo as simple as possible by neglecting local sources of turbulence, such as shear flow or cosmic-ray streaming (see \citet{Wiengarten-etal-2016} or \citet{Thomas-etal-2023}), and by choosing the K\'arm\'an-Taylor constants as $\alpha = 0.4$ and $\beta = \alpha/2=0.2$ \citep[see][and references therein]{Kleimann-etal-2023}.
Our choice to adopt the commonly used assumption of equipartition between kinetic and magnetic energy for the small-scale turbulent parts of $\vec{U}$ and $\vec{B}$, i.e., to assume
\begin{equation}
  \left< \delta \vec{b}^2 \right> =
  \left< \mu_0\, \rho\, \delta \vec{u}^2 \right>
  \ \Leftrightarrow \ \sigma_D = 0,
\end{equation}
is justified in Sect.~\ref{sec:sigma_D}.

\begin{figure*}
  \centering
  \includegraphics[width=0.92\linewidth]{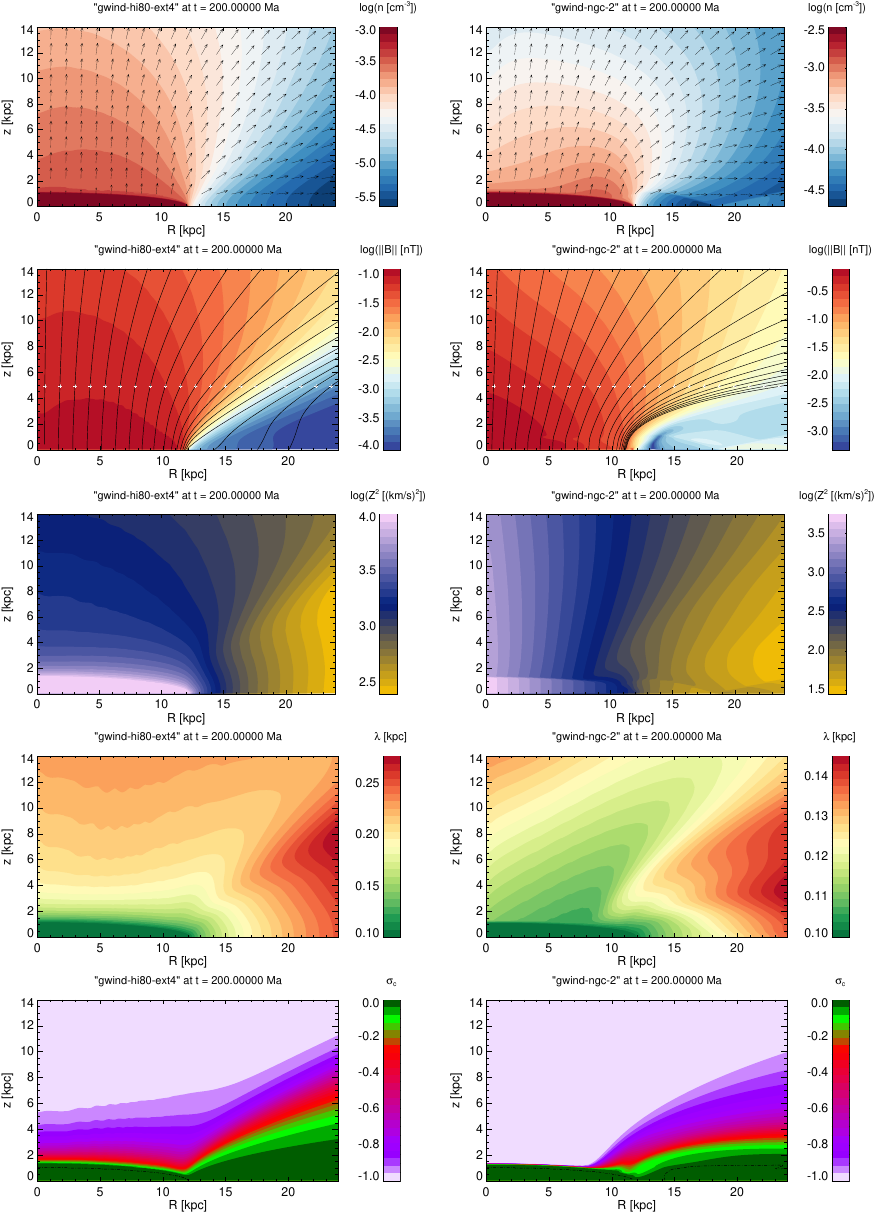}
  \caption{ Left: Poloidal ($\varphi=0$) cuts of the HI-ext run after convergence at $t=200$\,Ma. From top to bottom:
    $\log(n)$ with (poloidal) velocity arrows overplotted, $\log(\|\vec{B}\|)$ with field lines overplotted, and white crosses marking the field line seed points, $\log(Z^2)$, $\lambda$, and $\sigma_\mathrm{c}$.
    Right: The same quantities for the N4631 run. Note the different scalings used on both sides for all quantities except $\sigma_\mathrm{c}$. 
  }
  \label{fig:ext-vs-ngc}
\end{figure*}

The resulting steady-state situation is illustrated on the left side of Fig.~\ref{fig:ext-vs-ngc}. 
In the upper left panel, the (poloidal) velocity field and number density field are shown in the meridional plane. Their overall structure is consistent with expectations for the given geometry and boundary values, exhibiting an approximately homogeneous flow in $z$-direction and a stratified number density above the central plane, which gradually transitions to a more ``radial'' flow at larger $R$. Due to flow expansion, the number density 
monotonically decreases with increasing galactocentric distance.

The magnetic field exhibits the familiar ``X-shape'' configuration often observed in edge-on galaxies \citep[e.g.,][]{Krause-etal-2020, Stein-etal-2020, Stein_EA:2025} that forms naturally as a consequence of the field lines being frozen into the flow and dragged outward.
A conceptual model of this effect and its resulting observational signature are available in \citet{Henriksen:2022}.
Indeed, visually comparing the directions of field lines and velocity arrows is suggestive of field-aligned flow in all parts of the halo, except perhaps in the high-beta region located in the outer parts of the disk plane. As a result of this coupling, the inclination angle, $\arctan(B_z/B_R)$, 
decreases outbound, consistent with several analytical models \citep{Ferriere-Terral-2014, Kleimann-etal-2019, Unger_Farrar:2024}.
While the overall magnetic structure above the disk appears reasonable, the extremely low field strength prevailing in the wedge-shaped outer region likely results from the evacuation of the magnetic field beyond the disk edge by the outward-directed wind. This disappearing magnetic field cannot be replenished due to the strictly enforced no-flow condition within the disk.
To summarize, we find that the large-scale density, velocity, and magnetic fields produced in the simulations represent a plasma environment that adequately resembles a dynamical galactic halo.

The next three panels in the left column of Fig.~\ref{fig:ext-vs-ngc} depict the spatial dependence of the turbulence quantities prevailing in the model.
The overall structure of the fluctuation energy, $Z^2$, has an approximate horizontal stratification and is therefore similar to that of the number density above the central plane. Notably, this turns into an almost vertical stratification (with significantly lower values) toward the edge of the ellipsoid enclosing the galaxy. While the former behavior also appears in the correlation length $\lambda$, $Z^2$ shows the strongest decrease at intermediate latitudes. Lastly, the normalized cross helicity $\sigma_c$ gradually increases in magnitude with increasing height $z$ (from $0$ to almost $-1$). Its actual gradient depends mainly on the Alfv\'en Mach number, as discussed in Appendix~\ref{app:sigma}.

Direct simulation confirms that, upon the vertical extension of the grid from the upper half-space \mbox{$z \in [0,z_\mathrm{max}]$} to \mbox{$z \in [-z_\mathrm{max}, z_\mathrm{max}]$}, the expected spatial symmetries given by
\begin{equation}
  \label{eq:symmetry-signs}
  \left. \left( B_R, B_Z, Z^2,  \sigma_\mathrm{c}, \lambda \right) \right|_{z<0} =
  \left. \left(-B_R, B_Z, Z^2, -\sigma_\mathrm{c}, \lambda \right) \right|_{z>0}
\end{equation}
are indeed obtained. The correlation between the sign of $\sigma_\mathrm{c}$ and the polarity of $\vec{B}$ is addressed in Appendix~\ref{app:sigma}.

The results discussed in this subsection serve as a reference case illustrating the principal structure of a dynamical halo without tuning the model parameters to any real galaxy. This last step of adapting the simulations to observed galaxies is presented in Sect.~\ref{sec:applications}.

\subsection{Relation to the Wiengarten et al.\ model}

In their investigation of turbulence in the inner heliosphere, \citet{Wiengarten-etal-2015} numerically solved the set of equations~(\ref{eq:conti})--(\ref{eq:lam}), amended with a turbulence energy source term given by
\mbox{$S \propto U V_\mathrm{A} \exp(-L_\mathrm{cav}/r)$},
to model the influence of pickup-ions outside a Sun-centered, spherical cavity of size $L_\mathrm{cav} = 8$\,au. They presented their results as contour maps of all relevant quantities in the poloidal plane, covering a radial extent of $[0.3, 100]$\,au.
While some similarities between our results and those shown in their Fig.~2 can indeed be identified, the different physical setting --~characterized by a large computational domain relative to the extension of the inner boundary and a \citet{Parker:1958} spiral magnetic field that is predominantly azimuthal~--  severely limits the usefulness of a detailed comparison. We therefore restrict ourselves to noting that the preexisting numerical \citet{Wiengarten-etal-2015} setup nevertheless provided an invaluable starting point for the development of the present galactic model through a sequence of incremental changes, which significantly facilitated tests and consistency checks along the way.

\section{Application to a typical galactic halo}
\label{sec:applications}

\subsection{The reference galaxy}
\label{sec:refgal}

The parameters for the HI-ext run were chosen in order to highlight the general properties of MHD turbulence while keeping the HD aspects as close to the HI-base case as possible. In this subsection, by contrast, we investigate the changes associated with a more realistic astronomical object. For this purpose, our hypothetical reference galaxy was modeled after the low-mass, edge-on spiral galaxy \object{NGC~4631}, with corresponding parameters summarized in the last column (``N4631'') of Table~\ref{tab:parameters}.
By using the observed stellar optical intensity to trace the stellar mass, we determined the vertical extension of the central ellipsoid. We assumed the light profile to decay as \mbox{$I(z)=I(0)\, \exp(-z/H)$} with a global scale height of $H=450$\,pc and required that 90\% of the mass be contained within $|z|<z_\mathrm{c}$, yielding 
\mbox{$z_\mathrm{c} = H \ln(10) \approx 1$\,kpc}.

A potential complication arises from the fact that \object{NGC~4631} is located within a group of ten galaxies \citep[for further observational data of the \object{NGC~4631} group, see][]{Kourkchi_Tully:2017}, and that the combined halo of this group --~estimated to have a mass of $10^{12.1}\,M_\sun$ and to extend over a scale of $r_{200}=224$\,kpc \citep{Wang_EA:2023}~-- would once again place this galaxy in the HD-unstable regime. We therefore assumed that our reference galaxy is not a member of a larger group, resulting in a smaller, less massive halo.

Together with a concentration parameter $c_\mathrm{h}=8$, we obtained a scaling radius of $r_{200}=115$\,kpc by minimizing the average deviation in gravitational acceleration from the Hernquist potential given by \citet{Martinez-Delgado_EA:2015} for \object{NGC~4631}. The resulting rotational velocity of 150\,km/s at a distance of 8\,kpc compares favorably to the value of 145\,km/s cited in \citet{Wang_EA:2023} for the halo of the entire \object{NGC~4631} group.

The winds of galaxies are believed to be enhanced --~or even initiated in the first place~-- by starburst activity and/or cosmic-ray pressure, neither of which are included explicitly in our present model. The wind acceleration seen in our simulations, which at present is exclusively determined by the thermal pressure gradient working against gravity, therefore, likely underestimates the actual wind acceleration to be expected for galaxies such as \object{NGC~4631}. For this reason, we resorted to implicit heating through an artificially reduced adiabatic exponent, a method that has been used frequently in the context of solar wind modeling to account for the heating of the solar corona \citep[e.g.,][and references therein]{Lugaz_EA:2007, Jacobs_Poedts:2011, Shi_EA:2022}. The HI-base simulations indicate that the mean temperature of the steady-state halo is lower than the base temperature of the disk, which supplies the initial thermal energy, by approximately a factor of two. Therefore, we scaled the value given by \citet{Wang_EA:2023} accordingly. We note that this scaling behavior could to some extent be an artifact of the strong velocity gradient induced by the zero-flow disk boundary condition, which enforces a cooling effect through the conversion of thermal to kinetic energy at a considerable rate within a rather narrow transition layer. Thus, the ratio of halo-to-disk temperatures observed in actual galaxies may not be as low as our estimates.

All parameters are listed in the rightmost column of Table~\ref{tab:parameters} together with references to their respective sources. In particular, we used an average value of the total magnetic field strength found in a sample of late-type galaxies. This value agrees with equipartition estimates for \object{NGC~4631} \citep{Mora-Partiarroyo:2019_CR} and corresponds to a scale length for this field that falls within the range measured for a sample of comparable face-on galaxies.
Since the boundary values for $B$ and $Z^2$ decrease radially as
\begin{align}
  B &= B_\mathrm{disk} \exp(-R/L_B) , \\
  \label{eq:Zsq_exp}
  Z^2 &= Z^2_\mathrm{disk} \exp(-2 R/L_B) ,
\end{align}
the use of Eq.~(\ref{eq:delta_b2}) reveals that the relative strength of the turbulent magnetic field $\sqrt{\delta b^2}/B \approx 0.1$ remains constant at the disk (but generally not elsewhere).
We do not currently take the poloidal field reversals observed in \object{NGC~4631} \citep{Mora-Partiarroyo:2019} into consideration.

\subsection{Simulation results}
\label{sec:results}

The right side of Fig.~\ref{fig:ext-vs-ngc} displays the resulting steady state of the N4631 run in a side-by-side comparison to the one from the HI-ext run.
While the general structure of both cases is similar, a number of differences can be discerned.
First, the number density above the disk center has a higher negative gradient in the N4631 case compared to the HI-ext case. Second, the wedge-like geometry surrounding the radial axis in the $z=0$ plane is narrower, with more magnetic field lines connected to the disk and with a higher field strength within the wedge. Third, the gradient of the cross helicity is much stronger (as a consequence of a lower Alfv\'en Mach number; see Appendix~\ref{app:sigma}) in the N4631 case, such that its modulus assumes its maximum value almost throughout the entire halo, decreasing only toward the wedge-shaped region at the disk edge. 
Fourth, the principal difference with the HI-ext case is visible in the fluctuation
energy $Z^2$, which is vertically stratified throughout the halo above the disk in the N4631 case as a result of the nonconstant condition~(\ref{eq:Zsq_exp}) imposed on the inner boundary. The influence of the latter is also visible in the reduced horizontal stratification of the correlation length, which features very small (albeit positive) gradients of order $(\partial_z \lambda)|_{R=0} = 0.0017$ (N4631), compared to $0.0057$ for HI-ext. Notably, the structure of Eq.~(\ref{eq:lam}) suggests that a nonzero turbulent source term $S>0$ would actually diminish this gradient even further, possibly even to the extent of reversing its sign.
Finally, the different magnitudes of the variables in the two cases result from the different boundary conditions listed in Table~\ref{tab:parameters}.
The overall field line structure, while X-shaped in both runs, becomes somewhat parabolic toward the outer parts of the disk, compared to the less curved field seen in HI-ext.
A detailed quantitative analysis of the poloidal magnetic field and its potential agreement with existing analytical models, while certainly of interest, is beyond the scope of this paper, and will therefore be addressed in a forthcoming publication.

A first assessment of these results as a reasonable representation of \object{NGC~4631}'s halo can be obtained based on the large-scale quantities.
In particular, by integrating the mass flux across the top, bottom, and outer radial boundaries, we obtain a (hot gas) mass loss rate of $3.86\, M_\sun$/yr that is, in general, consistent with estimates for galactic winds \citep[e.g.,][]{Quataert-etal-2022} and, in particular, with the value $\sim 1.0\, M_\sun$/yr found for \object{NGC~4631} by \citet{Wang_EA:1995} within the error margins quoted therein. A second assessment, based on the diffusion coefficients derived from the small-scale quantities, is provided in Sect.~\ref{sec:diffuse} below.

\begin{figure}
  \centering
  \includegraphics[width=\linewidth]{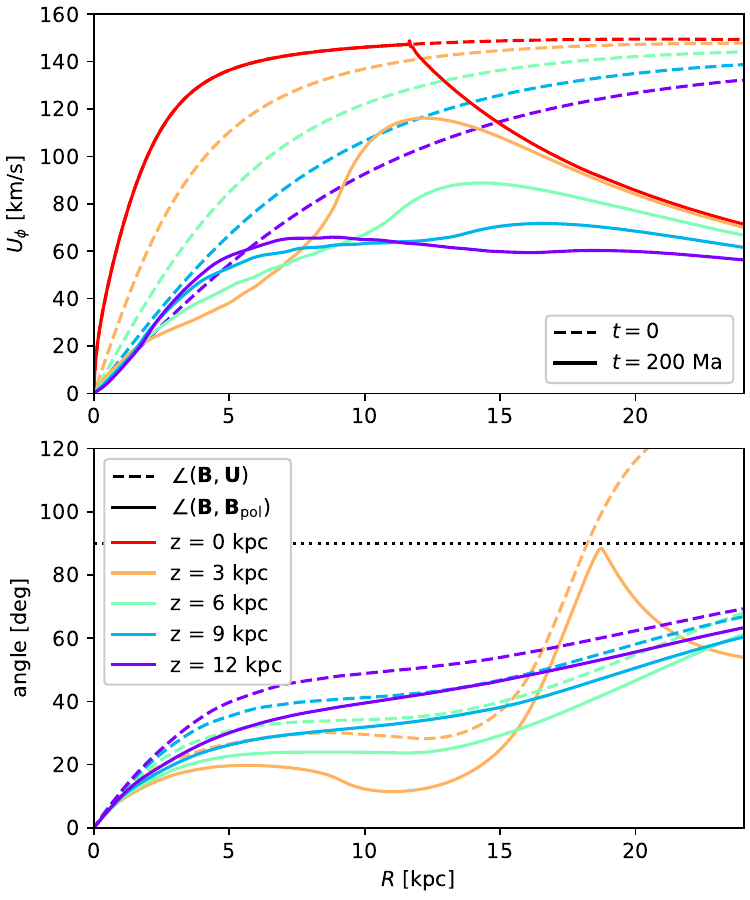}
  \caption{
    Azimuthal part of the flow and magnetic field in the N4631 run.
    Top: Cuts of $U_\varphi$ along selected heights $z$, as indicated by the legend in the bottom plot (shown only there for space). Dashed and solid lines represent the initial configuration and the final converged state, respectively.
    Bottom: Cuts of the angle between $\vec{B}$ and $\vec{U}$ (dashed), and between $\vec{B}$ and its projection onto the poloidal plane, for the same color-coded height levels as in the upper plot. The curve for $z=0$ is omitted due to symmetry.
    A horizontal (dotted) line at $90\degr$ is included for reference.}
  \label{fig:U-and-B}
\end{figure}

It is also insightful to consider the azimuthal components of the flow and the magnetic field, and their mutual relationship.
The top plot of Fig.~\ref{fig:U-and-B} shows the variation of $U_\varphi(R,z)|_{z=\mathrm{const}}$ with height above the plane. The dashed lines, which represent the expected flat galactic rotation curves as derived from Keplerian rotation via Eq.~(\ref{eq:v_Kepler}), diminish in magnitude with growing distance from the gravitating disk. In the final steady state, however, this magnitude is generally much lower --~except within the disk, where it is kept fixed~-- because conservation of a fluid element's angular momentum requires its azimuthal velocity to decay $\propto 1/R$. This is clearly seen, particularly for the $z=0$ curve.
Azimuthal velocity likewise decreases with height, at least at some distance from the $z$-axis. This is likely caused by ``magnetic braking,'' i.e., the process by which kinetic energy is lost as work done against magnetic tension when frozen-in field lines are forced to curve helically around the axis of rotation. Such a velocity lag in a galactic wind was also modeled analytically by \citet{Henriksen_Irwin:2016}, though they assumed field-aligned flow and zero rotation at the upper boundary.

\subsection{A parametric study of \texorpdfstring{$\sigma_D$}{sigmaD}}
\label{sec:sigma_D}

The normalized residual energy (also known as energy difference) $\sigma_D$ is difficult to constrain. For the outer heliosphere, models that approximate this quantity as a constant
\citep[e.g.,][]{Zank-etal-1996, Yokoi-etal-08, Breech-etal-2008} often assume
\mbox{$\langle \delta \vec{b}^2 \rangle = 2\langle \mu_0\, \rho\, \delta \vec{u}^2 \rangle$} (and thus imply $\sigma_D = -1/3$)
based on corresponding Voyager observations \citep{Matthaeus-Goldstein-1982}; 
see also the more recent analysis of Ulysses data by \citet{Perri_Balogh:2010}.
However, even in that context, the usefulness of this assumption was questioned by \citet{Kleimann-etal-2023} through simulations in which $D \equiv \sigma_D Z^2$ was treated as a dynamic variable. Therein, $\sigma_D$ was found to vary significantly, from positive values in the local interstellar medium to low negative values ($\approx -0.7$) just outside the solar termination shock. Since our present model approximates $\sigma_D$ as a constant, and in the absence of reliable observational galactic constraints, we find it useful to conduct a parameter study to establish the effect and sensitivity with which turbulence-related variables respond to changes in $\sigma_D \in [-1,+1]$ while keeping all other parameters equal to those of HI-ext. 

\begin{figure*}
  \centering
  \includegraphics[width=\linewidth]{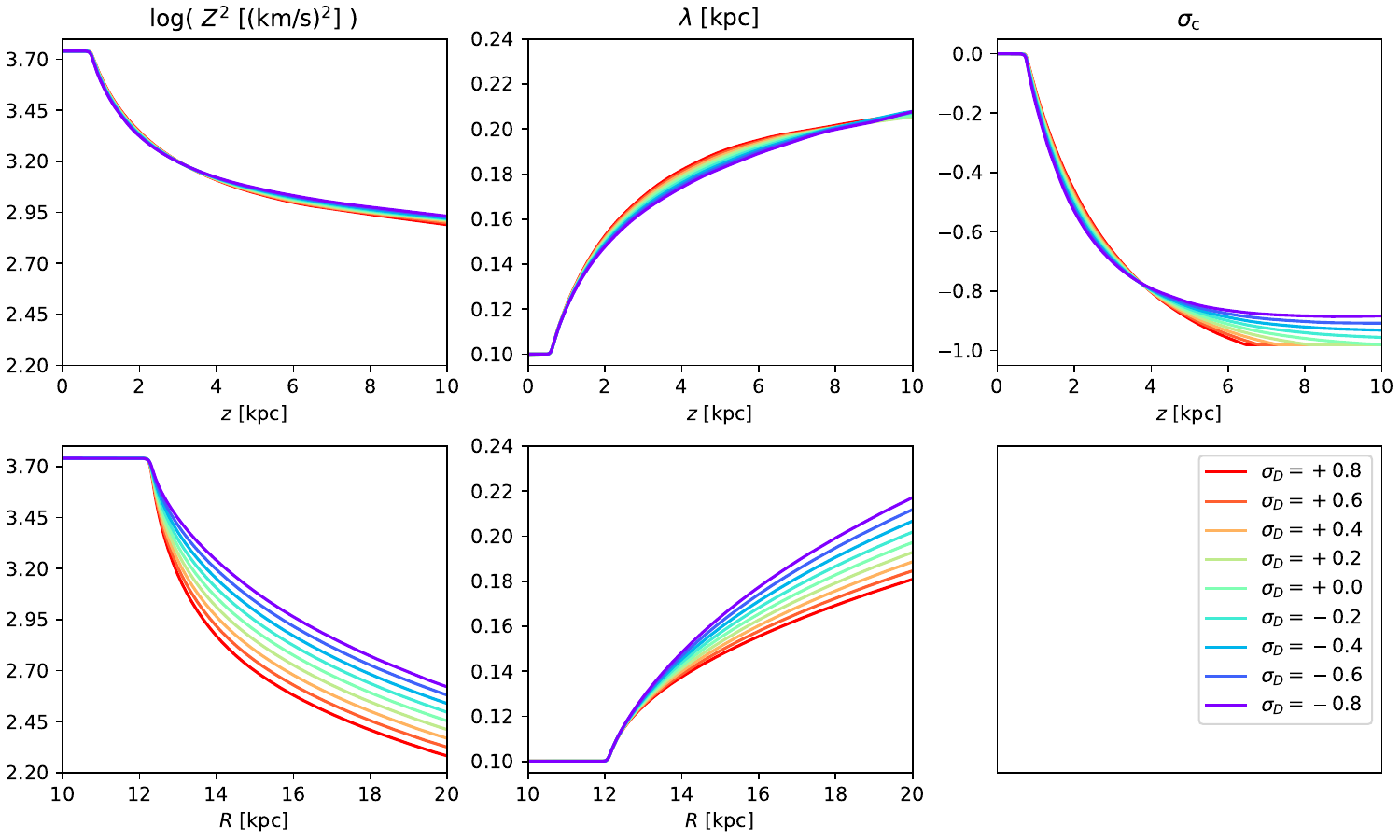}
  \caption{\label{fig:sigmaD_runs}
    Profiles of $\log(Z^2)$ (left), $\lambda$ (middle), and $\sigma_\mathrm{c}$ (right) along straight lines through the halo.
    Top: Cutting along $z$ at a fixed axial distance of $R=10$\,kpc.
    Bottom: Cutting along $R$ at a fixed height of $z=0$.
    The plot for $\sigma_\mathrm{c}(R)$ is omitted, as $\sigma_\mathrm{c}=0$ at $z=0$ due to  symmetry.}
\end{figure*}

Figure~\ref{fig:sigmaD_runs} illustrates the resulting variation through one-dimensional profiles of $Z^2$, $\lambda$, and $\sigma_\mathrm{c}$, both as a function of height at a specific distance of $R=10$\,kpc and as a function of radial distance in the disk plane.
As seen in the upper row of the plots, both $Z^2$ and $\lambda$ show minimal variation with height, while $\sigma_\mathrm{c}$ exhibits a nontrivial degree of variability at larger $|z|$.
The artificial numerical clamping required at larger values of $\sigma_D$ --~to ensure $\sigma_\mathrm{c} \ge -1$ (or, for technical reasons, $\ge -0.98$) and prevent $f^\pm(\sigma_\mathrm{c})$ from becoming ill-defined~-- is also readily apparent. We note that in an actual application, Eq.~(\ref{eq:sigma_circle}) would restrict the range of admissible $\sigma_\mathrm{c}$ values even further.

Generally, the spread seen in all three quantities for different $\sigma_D$ increases with radial distance $R$ (not shown) and becomes quite small (though not zero) at $R=0$. The fact that all observed variations with $\sigma_D$ are 
monotonic and approximately proportional to $\sigma_D$, together with their limited extent of variability, justifies our choice of the intermediate case $\sigma_D=0$ for both the HI-ext and the N4631 run. An additional advantage of this choice is that it eliminates the need for corrective interventions to ensure that Eq.~(\ref{eq:sigma_circle}) is satisfied; in this case, it simply reads $|\sigma_\mathrm{c}| \le 1$. Furthermore, the dependencies displayed in Fig.~\ref{fig:sigmaD_runs} can be used to qualitatively estimate how the two-dimensional contours of $Z^2$, $\lambda$, and $\sigma_\mathrm{c}$ in Fig.~\ref{fig:ext-vs-ngc} would change if a nonzero $\sigma_D$ were used.

The variation observed outside the disk at $z=0$ (i.e., the bottom row of plots in Fig.~\ref{fig:sigmaD_runs}) is larger than in the perpendicular direction. However, as discussed in Sect.~\ref{sec:hi_ext} above, the physical interpretation of MHD quantities that can be extracted from the simulations in this region is likely affected by the static flow conditions enforced within the disk. Therefore, for analyzing the turbulent properties and their dependence on $\sigma_D$, the top row of plots in Fig.~\ref{fig:sigmaD_runs} should be preferred over the bottom one.

\subsection{The spatial diffusion tensor in the halo}
\label{sec:diffuse}

Once the amplitude of the magnetic fluctuations is known, the elements of the spatial diffusion tensor of cosmic rays can be computed. Among the various theories describing the anisotropic diffusion process \citep[e.g.,][]{Shalchi-2021}, we selected a nonlinear state-of-the-art formulation in order to illustrate how the diffusion tensor can be computed in an ab~initio manner for the entire halo of a given galaxy. 

The general form of the spatial diffusion tensor in a local field-aligned coordinate system is given by
\begin{equation}
  \tens{\kappa} = \left(
  \begin{matrix}
    \kappa_\perp & 0 & 0 \\
    0 & \kappa_\perp & 0 \\
    0 & 0 & \kappa_\|    \\
  \end{matrix}
  \right)
\end{equation}
\citep[e.g.,][]{Jokipii-1971, Effenberger-etal-2012, Snodin-etal-2016}, where $\kappa_{\|,\perp}$ denote the diffusion coefficients parallel and perpendicular to the local magnetic field $\vec{B}$. Off-diagonal elements can arise from nonaxisymmetric turbulence \citep[e.g.,][]{Weinhorst-etal-2008} or particle drifts \citep[e.g.,][]{Kota-Jokipii-1983}, neither of which is considered here.

The dependence of the diffusion coefficients on magnetic turbulence has been studied analytically and numerically over many years \citep[e.g.,][]{Schlickeiser-2002, Shalchi-2009}. The studies have shown that parallel diffusion can be treated satisfactorily within the framework of quasi-linear theory. Accordingly, we used the following relation for the corresponding coefficient \citep[e.g.,][]{Moloto-Engelbrecht-2020}: 
\begin{equation}
  \label{eq:kappa_para}
  \kappa_{\parallel}
  = \frac{v \, k_\mathrm{m} \, r_\mathrm{g}^2 }{\pi} \frac{s}{s-1}
  \frac{B^2}{\delta b_\mathrm{sl}^2}
  \left[ \frac{1}{4} + \frac{2 (k_\mathrm{m} \, r_\mathrm{g})^{-s}}{(2-s)(4-s)}\right],
\end{equation}
where $s=5/3$ is the inertial range spectral index for the magnetic energy spectrum,
$k_\mathrm{m} = 1/\lambda_\mathrm{sl}$ is the lowest wave number of the inertial range of the slab spectrum (i.e., corresponding to the slab correlation scale), and $v$ and $r_\mathrm{g}$ are the particle speed and gyro radius, respectively. Given the uncertainties regarding the turbulence properties in a galactic halo, we employed the ``heliospheric'' result $\lambda_\mathrm{sl}\gtrsim\lambda_\mathrm{2d} = \lambda$, where $\lambda$ is the solution to Eq.~(\ref{eq:lam}).

The various theories yield differing results for the perpendicular diffusion coefficient \citep[e.g.,][]{leRoux-etal-1999, Florinski-Pogorelov-2009, Shalchi-2021}. 
In view of the uncertainty regarding the nature of the turbulence in a galactic halo, we considered a commonly used approach, namely that resulting from the so-called nonlinear guiding center theory \citep[NLGC,][]{Matthaeus-etal-2003, Bieber-etal-2004, Shalchi-etal-2004}:
\begin{equation}
  \label{eq:kappa_perp}
  \kappa_{\perp} 
  = \left[v \sqrt{\frac{\pi}{3}}\frac{2s-2}{s} 
    \frac{\Gamma(s/2)}{\Gamma(s/2-1/2)}\lambda_\mathrm{2d}
    \frac{\delta b_\mathrm{2d}^2}{B^2}\right]^{2/3} \kappa_{\|}^{1/3},
\end{equation}
where $\Gamma(x)$ is the gamma function and $\lambda_\mathrm{2d} = \lambda$ is the correlation scale of the quasi-two-dimensional fluctuations, i.e., the solution of Eq.~(\ref{eq:lam}).
For a 10\,GeV proton, gyration radii in the halo are typically of order
\begin{equation}
  r_\mathrm{g} = \frac{\gamma \, m_\mathrm{p} \, v}{e \, B}
  \sim \frac{10\, \mathrm{GV}/c}{0.5\, \mathrm{nT}}
  \approx 6 \times 10^{10}\, \mathrm{m} \approx 2 \times 10^{-6}\, \mathrm{pc} ,
\end{equation}
where $e$ denotes the elementary charge and $c$ the speed of light.
The NLGC has successfully been tested with full-orbit simulations \citep[e.g.,][]{Dosch-etal-2013} and has been applied to various astrophysical scenarios, reaching from the heliosphere \citep[e.g.,][]{Moloto-Engelbrecht-2020} via supernova shock acceleration \citep{Li-etal-2009} to the interstellar transport of cosmic rays \citep{Shalchi-etal-2010}.

\begin{figure}
  \centering
  \includegraphics[width=\linewidth]{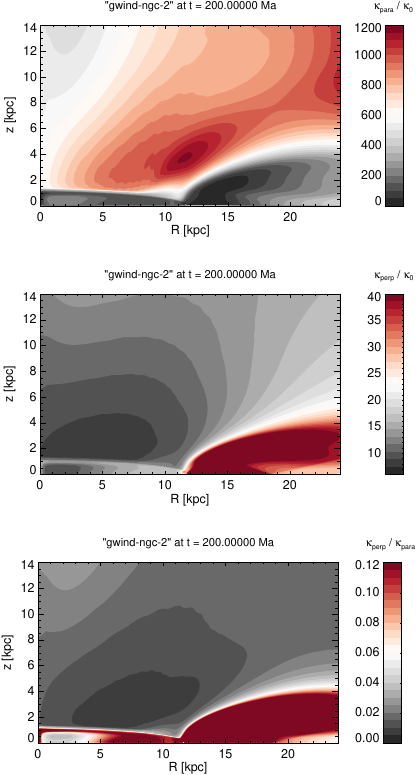}
  \caption{
    Diffusion coefficients in the halo in units of $\kappa_0=10^{25}$\,m$^2$/s,
    derived for 10\,GeV protons using data from the N4631 run.
    Top: parallel diffusion $\kappa_\parallel/\kappa_0$.
    Middle: perpendicular diffusion $\kappa_\perp/\kappa_0$.
    Bottom: anisotropy ratio $\kappa_\perp/\kappa_\parallel$ according to Eq.~(\ref{eq:kappa_ratio}). Note that the color scale in the middle and bottom panels have been truncated at their upper ends to allow for more fine structure to be visible above the disk.}
  \label{fig:kappa}
\end{figure}

For an illustration of the resulting diffusion coefficients in a galactic halo, 
we used $\delta b_\mathrm{2d}^2 = \delta b^2$ (see Sect.~\ref{sec:smallscale}) and a constant ratio $\delta b_\mathrm{sl}^2/\delta b_\mathrm{2d}^2 = 0.25$, well within the range of the values used in two-component models \citep[e.g.,][]{Adhikari-etal-2017} and derived from observations \citep[e.g.,][]{Bieber-etal-1996}. Furthermore, we made use of the fact that, for the halo environment, the expression inside the square brackets of Eq.~(\ref{eq:kappa_para}) is entirely dominated by the second term, implying
\begin{align}
  \kappa_\parallel &\approx 8.185\, v\, \left(\lambda^2 \, r_\mathrm{g} \right)^{1/3}
  \left(\frac{\delta b^2}{B^2}\right)^{-1} , \\
  \kappa_\perp &\approx 0.991\, v\, \left(\lambda^8 \, r_\mathrm{g} \right)^{1/9}
  \left(\frac{\delta b^2}{B^2}\right)^{1/3} ,
\end{align}
and hence an anisotropy of
\begin{equation}
  \label{eq:kappa_ratio}
  \frac{\kappa_\perp}{\kappa_\parallel}
  \approx 0.121 \left( \lambda / r_\mathrm{g} \right)^{2/9}
  \left(\frac{\delta b^2}{B^2}\right)^{4/3} .
\end{equation}
We note that in this notation the $\delta b^2$ is related to the simulated $Z^2$ via Eq.~(\ref{eq:delta_b2}).
Figure~\ref{fig:kappa} shows contours of $\kappa_\parallel$ and $\kappa_\perp$ for the N4631 run, together with their ratio~(\ref{eq:kappa_ratio}), evaluated for a proton energy of 10\,GeV as an example. 
Together with the turbulence quantities computed for \object{NGC~4631}, the above model for the diffusion tensor yields values within the expected range: the parallel diffusion coefficient is of the same order as found, for example, by \citet{Thomas-etal-2023}, while the perpendicular coefficient is, as expected, significantly lower. The former varies across the halo by approximately a factor of 10, with highest values at mid-latitudes, and comparatively low values above the central plane and, in particular, within the wedge at the edge of the disk. In contrast, the perpendicular diffusion coefficient assumes its highest values inside that wedge and has the lowest values above the disk below a halo height of 3-4~kpc. Such ``inverse'' behavior is, of course, to be expected from the fact that $\kappa_{\|} \propto ({\delta b^2}/{B^2})^{-1}$ and $\kappa_\perp \propto ({\delta b^2}/{B^2})^{1/3}$.

While these diffusion coefficients depend on both large-scale (i.e., mass density and magnetic field) and small-scale quantities (i.e., fluctuation energy and correlation length), their sensitivity to these quantities differs;
see Eqs.~(\ref{eq:delta_b2}), (\ref{eq:kappa_para}), and (\ref{eq:kappa_perp}).
In particular, the correlation length does not vary strongly (by less than a factor of two) and, by consequence, only weakly influences the magnitude of the diffusion. The other three quantities, however, vary by more than a factor of 100, and therefore dominate the coefficients' behavior. The geometry of the region of lowest and highest values of the parallel and perpendicular diffusion coefficient clearly reflects the influence of the magnitude of the large-scale magnetic field, for which the associated wedge structure is most clearly pronounced (see the right panel in the second row of Fig.~\ref{fig:ext-vs-ngc}).
The variation outside of this region results from the interplay between variations in the magnetic field, the density, and the fluctuation energy. The contour patterns
in the first three panels of the right column of Fig.~\ref{fig:ext-vs-ngc} suggests that the variation of the parallel diffusion coefficient, exhibiting a ``ridge'' of high values at mid-latitudes, is mostly influenced by the fluctuation energy, which has low values along the wedge's boundary and increases toward the halo above the disk. This trend is, of course, ``modulated'' by the gradients of the density and magnetic field. The inverse and weaker variation of the perpendicular diffusion coefficient results from the exponent $1/3$ instead of $-1$ in Eqs.~(\ref{eq:kappa_perp}) and (\ref{eq:kappa_para}), respectively.

Generating maps of the diffusion coefficients for other cosmic-ray energies is straightforward. Since the energy dependence of the coefficients is prescribed by the diffusion model and does not depend on the simulations above, the results will be similar to those shown in Fig.~\ref{fig:kappa}, but scaled according to their dependence on the gyro radius. Therefore, the structure of the diffusion tensor of cosmic rays in a galactic halo like that of \object{NGC~4631} is well illustrated by the results shown in that figure. A clear finding is that the diffusion coefficients and the anisotropy of the diffusion are far from uniform in such a halo, which adds to the complexity of modeling cosmic-rays transport in galactic halos.

\section{Summary and conclusions}
\label{sec:summary}

Through this study, to the best of our knowledge, we established for the first time the evolution of MHD turbulence in galactic halos. To this end, we created as a first step a simple reference model for a ``generic,'' magnetized, turbulent galactic wind. The setup of this model was preceded by thorough testing of the numerical implementation. Notably, the purely hydrodynamic, nonturbulent case, which was tested against the classical simulations conducted by \citetalias{Habe-Ikeuchi-1980}, revealed that for a sufficiently massive galaxy (consisting of disk and dark-matter halo), the outflow is unstable above the central disk, regardless of galactic rotation. While this instability disappears with sufficiently low gravitating mass, it is nonetheless likely to be highly relevant for more massive galaxies such as the Milky Way.

In the generic case, the large-scale field variables (i.e., mass density, velocity, temperature, and magnetic field) provided a suitable background for the small-scale quantities characterizing the turbulence (fluctuation energy, correlation length, and cross helicity) to evolve self-consistently.
While the former correspond to general expectations for galactic winds, the spatial distributions of the latter represent new results that illustrate their principal variation in galactic halos.

In a third step, we simulated, with some approximation, the dynamical halo of the galaxy \object{NGC~4631}. As a fourth and final step, following a discussion of the differences in the generic case, we computed the diffusion tensor of cosmic rays in the halo of that galaxy by employing a nonlinear state-of-the-art theory of anisotropic spatial diffusion. The results not only yield values for the diffusion coefficients in the expected range, but also reveal that these coefficients, along with the anisotropy of the diffusion, show significant spatial variation in a galactic halo. These findings add to the complexity of modeling the transport of cosmic rays in galactic halos.

Although the present model establishes a framework for studying the turbulence in the dynamical halos of galaxies, several extensions are clearly necessary --~if not essential~-- to enhance the comparability of simulation results with observations.
First, the turbulence model should be extended from a one-component to a two-component model, in which the slab and two-dimensional fluctuations are treated separately.
Second, the multiphase nature of the thermal plasma could be addressed by using multiple localized, time-dependent outflows from a galactic disk. This would enable us to distinguish between different phases, such as warm and hot gas.
Third, the potential instability of galactic outflows in more massive galaxies should be studied, including its consequences for partial (as opposed to global) winds, for the evolution of turbulence, and for its impact on cosmic-ray transport in galactic halos. Last but not least, cosmic-ray feedback on the thermal plasma should be modeled explicitly by incorporating a corresponding transport equation into the model.

\section*{Data availability}

The \textbf{ancillary movie file} associated to Fig.~\ref{fig:quad_Uz} is available at
\href{https://www.aanda.org/articles/aa/olm/2025/07/aa53873-25/aa53873-25.html}{https://www.aanda.org}.

\begin{acknowledgements}
  We thank the anonymous referee for their helpful comments. Furthermore, we gratefully acknowledge financial support by the German Research Foundation (Deutsche Forschungsgemeinschaft, DFG) through the Collaborative Research Center (CRC; i.e., Sonderforschungsbereich, SFB) 1491.
\end{acknowledgements}

\bibliographystyle{aa}
\bibliography{dynhalo_lit}

\begin{appendix}
  
  \section{The hyperbolic magnetic field}
  \label{app:hyper_B}
  
  For elliptic coordinates $(\mu, \nu) \in \mathbb{R}_{\ge 0} \times [0,2\pi)$ defined as usual through
    \begin{align}
      \label{eq:ellipt_R}
      R &=a\ \cosh \mu \ \cos \nu , \\
      \label{eq:ellipt_z}
      z &=a\ \sinh \mu \ \sin \nu ,
    \end{align}
    we wish to derive a magnetic field with its field lines on hyperbolas (lines of constant $\nu$) because they are normal to ellipses (lines of constant $\mu$) and in particular to the contour of our central ellipsoid with focal distance
    $a = \sqrt{R_\mathrm{c}^2 - z_\mathrm{c}^2}$.
    Eliminating $\mu$ from Eqs.~(\ref{eq:ellipt_R}) and (\ref{eq:ellipt_z}), we obtain
    \begin{align}
      \label{eq:label_C}
      2 \, C(R,z) :=&\ r^2-\sqrt{(a^2+r^2)^2- (2 a R)^2} \nonumber \\
      =&\ a^2 \cos (2 \nu)
    \end{align}
    (where $r^2 = R^2 + z^2$) as an alternative field line label that, for the present purpose, is more convenient than $\nu$ itself. (We note in passing that inverting the sign in front of the square root in Eq.~(\ref{eq:label_C}) would make the expression equivalent to $a^2 \cosh(2\mu)$, which could thus be used in place of $\mu$ as an alternative label for confocal ellipses.)
    Being constant along field lines by construction, we can not only use $C(R,z)$ to easily draw field lines (as contours of constant label $C$) but also as a flux function to obtain the desired divergence-free magnetic field via a vector potential \mbox{$\vec{A} = (C/R) \, \vec{e}_\varphi$} as
    \begin{equation}
      \label{eq:B_hypfromA}
      \tilde{\vec{B}} = \nabla \times \vec{A}
      = \frac{1}{R} \left(
      -\frac{\partial C}{\partial z} \, \vec{e}_R
      +\frac{\partial C}{\partial R} \, \vec{e}_z
      \right) .
    \end{equation}
    Explicit evaluation of this equation leads to the intermediate field
    \begin{align}
      \tilde{B}_R &= \left( \frac{a^2+r^2 }{\sqrt{(a^2+r^2)^2-(2 a R)^2 }} -1 \right) \frac{z}{R} , \\
      \tilde{B}_z &= \left( \frac{a^2-r^2 }{\sqrt{(a^2+r^2)^2-(2 a R)^2 }} +1 \right) ,
    \end{align}
    which already has the correct geometrical properties. 
    Now let $b(R)$ be the desired total field strength at radius $R\in  [0,R_\mathrm{c}]$ on the ellipsoid's surface. The correct magnetic field components at position $(R,z)$ are obtained through a normalization and re-scaling of $\tilde{\vec{B}}$ via
    \begin{equation}
      \vec{B}(R,z) = \frac{b(R_0)}{\|\tilde{\vec{B}}(R_0,z_0)\|} \, \tilde{\vec{B}}(R,z) ,
    \end{equation}
    where $(R_0,z_0)$ is the position at which the field line passing through $(R,z)$ intersects the ellipsoid, and can therefore be obtained through the conditions $C(R_0,z_0)=C(R,z)$ and
    \begin{equation}
      r_\mathrm{e}(R_0, z_0)^2 \equiv (R_0/R_\mathrm{c})^2 + (z_0/z_\mathrm{c})^2 = 1 
    \end{equation}
    as
    \begin{equation}
      \frac{R_0}{R_\mathrm{c}} = \sqrt{\frac{1}{2} + \frac{C(R,z)}{a^2}}
      \quad \mbox{and} \quad
      \frac{z_0}{z_\mathrm{c}} = \sqrt{\frac{1}{2} - \frac{C(R,z)}{a^2}} ,
    \end{equation}
    respectively. In the spherical limit
    ($z_\mathrm{c}/R_\mathrm{c} \rightarrow 1$), we recover the radial field
    \begin{equation}
      \vec{B}|_{a \rightarrow 0} = b \left( \frac{R \, R_\mathrm{c}}{r} \right)
      \, \left(\frac{R_\mathrm{c}}{r}\right)^2
      \left( \frac{z}{|z|} \frac{R}{r} \, \vec{e}_R + \frac{|z|}{r} \, \vec{e}_z \right)
    \end{equation}
    whose polarity changes across the $(z=0)$ plane.
    
    \section{More details on the inner boundary}
    \label{app:boundary}
    
    The fact that the ellipsoidal boundary does not coincide with the coordinate surfaces of the cylindrical grid makes this setup susceptible to numerical artifacts, which typically take the form of curved rays emanating from ``steps'' in the staircase-like structure that separates cells inside the boundary from those on the outside. The problem is mitigated through a smoothing procedure
    \begin{equation}
      q(\vec{r},t) \leftarrow \,
      f(\vec{r}) \, q(\vec{r},t) + [1-f(\vec{r})] \, q_0(\vec{r})
    \end{equation}
    replacing the quantity $q$ at cell position $\vec{r}$ and time(step) $t$ by a spatially weighted average of itself and $q_0$, the corresponding constant boundary value.
    Therein, $f$ transitions smoothly from $0$ to $1$ within a boundary layer of thickness $2\, h$ centered on the actual ellipsoid's surface. More specifically, we first define
    \begin{equation}
      r_\mathrm{b}(R,z;w) = \sqrt{\left(\frac{R}{R_\mathrm{c}+w}\right)^2+
        \left(\frac{z}{z_\mathrm{c}+w}\right)^2}
    \end{equation}
    generalizing Eq.~(\ref{eq:r_ell}) such that $r_\mathrm{e}(R,z) = r_\mathrm{b}(R,z;0)$.
    Then, the weighting function $f(\vec{r})$ is defined as
    \begin{equation}
      f(R,z) = \left\{ \begin{array}{r@{ \ \ : \ \ }l}
        0 & r_\mathrm{b}(R,z;w_0-h) < 1 \\
        1 & r_\mathrm{b}(R,z;w_0+h) > 1 \\
        (3 -2 u) u^2 & \mbox{else}
      \end{array} \right.
    \end{equation}
    with the shorthand
    \begin{equation}
      u = \frac{r_\mathrm{b}(R,z;w_0-h)
        \left[ 1-r_\mathrm{b}(R,z;w_0+h) \right]}{
        r_\mathrm{b}(R,z;w_0-h) - r_\mathrm{b}(R,z;w_0+h) } ,
    \end{equation}
    establishing a smooth transition in the intermediate region. A suitable value of $h=0.16$\,kpc (equal to four cell spacings) was found on a trial-and-error basis.
    The position of the relevant boundary is given by $w_0=0$ (i.e., the actual galactic ellipsoid) except for the turbulence quantities $Z^2$ and $H_\mathrm{c}$, for which it is is enlarged to $w_0=0.2$\,kpc.
    The reason is that, as can be seen from Eq.~(\ref{eq:cross}), the behavior of $\sigma_\mathrm{c} \equiv H_\mathrm{c}/Z^2$ crucially depends on the gradient of $\vec{U}$ (which attains unphysically high values near the boundary due to our strict zero-flow boundary condition) as well as on the magnitude of $\vec{U}$ itself (which is too low at $w_0=0$ for the same reason). At a distance of $0.2$\,kpc away from the ellipsoid, however, the wind was accelerated to a speed sufficiently close to its ``true'' value, allowing it to affect the turbulence.
    
    \section{Possible reasons for the hydrodynamic stability seen in
      \citetalias{Habe-Ikeuchi-1980}}
    \label{app:hi80-steady}
    
\begin{figure}
      \centering
      \includegraphics[width=\linewidth]{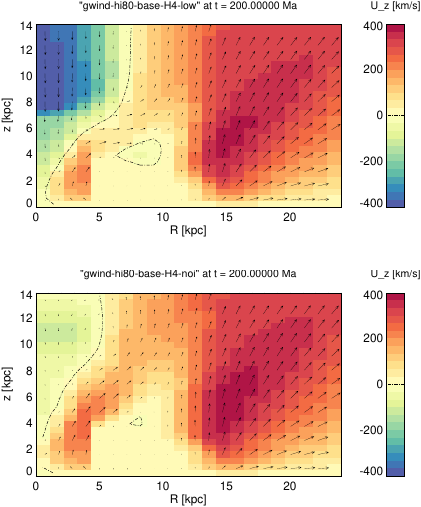}
      \caption{Contours of $U_z$ and poloidal velocity vectors for the HI-base run when using a grid size equal to that of \citetalias{Habe-Ikeuchi-1980}, with colored blocks indicating actual computational cells.
        Top: with zero-gradient boundary conditions at all outer boundaries.
        Bottom: additionally enforcing a no-inflow condition at $z=z_\mathrm{max}$. The length of velocity arrows is directly proportional to $\sqrt{U_R^2 + U_z^2}$ in order to ease comparison with Fig.~3 of \citetalias{Habe-Ikeuchi-1980}.}
      \label{fig:hi80-coarse}
    \end{figure}

    One of the few remaining differences between the original ``wind-type'' simulation of \citetalias{Habe-Ikeuchi-1980} and our replication thereof discussed in Sect.~\ref{sec:habe_ikeuchi} is the much coarser computational grid ($N_R \times N_z = 22 \times 30$) of the former. In order to ascertain whether the high numerical diffusivity associated with such a coarse (by today's standards) resolution prevented the instability from arising, we also replicated the HI-base run with the original resolution, and at $t=200$\,Ma obtained the flow field presented in the top panel of Fig.~\ref{fig:hi80-coarse}. Evidently, the inflow is already very pronounced, unlike the corresponding situation depicted in Fig.~3 of \citetalias{Habe-Ikeuchi-1980}, which shows essentially no near-axis fluid motion. As the simulation progresses, the inflow continuously grows in both magnitude and spatial extent, similarly to the nonrotating case shown in Fig.~\ref{fig:quad_Uz}c but covering a wider radial domain. The complete absence of any Kelvin-Helmholtz-like flow patterns can now unambiguously be traced to the reduced resolution, since this is the sole difference between the runs depicted in Figs.~\ref{fig:quad_Uz}a and \ref{fig:hi80-coarse}.
    Interestingly, we eventually succeeded in obtaining a steady-state wind exhibiting considerable similarity to the original work by implementing a zero-inflow condition at the upper boundary $z=z_\mathrm{max}$. 
    This finding strongly suggest that a similar mechanism had been implemented by \citetalias{Habe-Ikeuchi-1980}, although a definitive answer is precluded by the fact that these authors unfortunately did not discuss the exact nature of their employed boundary conditions.
    
    \section{A one-dimensional model of the halo's cross helicity}
    \label{app:sigma}
    
    To better understand the very different behavior of $\sigma_\mathrm{c}$ exhibited in the two cases shown in the bottom row of Fig.~\ref{fig:ext-vs-ngc}, consider Eqs.~(\ref{eq:Z2}) and (\ref{eq:cross}) evaluated for the special case of field-aligned flow ($\vec{u} \parallel \vec{B}$) within a flux tube whose cross section changes according to $A(s)=A_1 \, s^\delta$, where $s$ is a suitably normalized coordinate along the tube's axis, and a subscript ``1'' indicates evaluation at $s=1$. The divergence  of a vector $\vec{X} \in \{ \vec{U}, \vec{B} \}$, which then only has a single component (in $s$-direction), is in this case computed according to
    \begin{equation}
      \nabla \cdot \vec{X} = \frac{1}{s^\delta} \frac{\partial}{\partial s}
      \left( s^\delta \, X_s \right) .
    \end{equation}
    For constant speed $U$, conservation of mass and magnetic flux implies that both $\rho$ and $B=B_s$ are proportional to $s^{-\delta}$, hence
    $V_\mathrm{A} \propto s^{-\delta/2}$ and thus
    \begin{equation}
      \nabla \cdot \vec{V}_\mathrm{\!\!A} = \frac{\delta}{2} \, \frac{V_\mathrm{A,1}}{s^{1+\delta/2}}
      \quad \mbox{and} \quad \nabla \cdot \vec{U} = \delta \, \frac{U}{s} ,
    \end{equation}
    where $V_\mathrm{A,1} = V_\mathrm{A}|_{s=1}$.
    It can be shown that, for $\sigma_D=0$ and in the absence of any $\alpha$-related source terms, the coupled system of  Eqs.~(\ref{eq:Z2}) and (\ref{eq:cross}) admits the exact stationary solution
    \begin{align}
      \label{eq:sig_1d}
      \sigma_\mathrm{c}(s) &=
      \frac{ 2 m \left(m^2 s^{\delta/2}-1 \right) \left(s^{\delta/2}-1\right) }{
        4 m^2 \, s^{\delta/2} - (m^2+1) \left(m^2 \, s^\delta +1 \right) } , \\
      \label{eq:Zsq_1d}
      \frac{Z^2(s)}{Z^2(1)} &=
      \frac{(m^2+1) \left(m^2 \, s^\delta + 1\right) \, s^{\delta/2}-4 \, m^2 \, s^\delta}{\left(m^2 \, s^\delta-1 \right)^2}
    \end{align}
    for the boundary condition $\sigma_\mathrm{c}(1)=0$, where $m:= U/V_\mathrm{A,1}$ is the Alfv\'en Mach number at $s=1$.
    For the limiting case of nearly straight flux tubes ($\delta=0$) near the galaxy's rotational axis, we obtain constant $\sigma_\mathrm{c}(s)=0$ and $Z^2(s)=Z^2(1)$ as expected, while both decrease monotonically for any $\delta>0$.
    
\begin{figure}
      \centering
      \includegraphics[width=\linewidth]{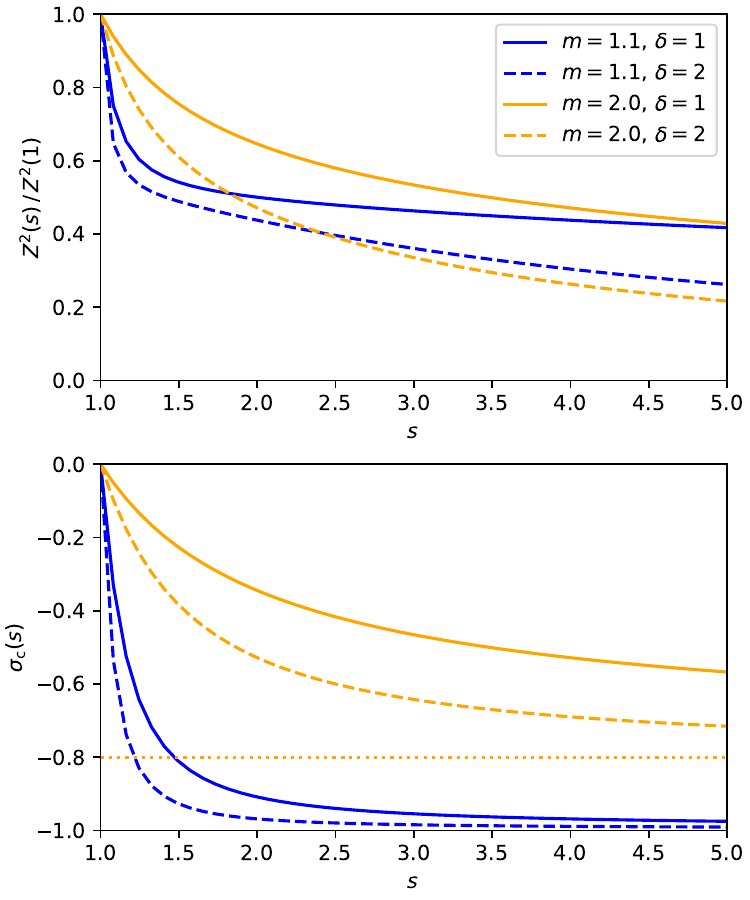}
      \caption{ Top: Normalized $Z^2(s)$ according to Eq.~(\ref{eq:Zsq_1d}) for the simplified one-dimensional case, plotted for selected values of Alfv\'enic Mach number $m \in \{1.1, 2.0\}$ and expansion exponent $\delta \in \{1, 2\}$.
        Bottom: The same for $\sigma_\mathrm{c}(s)$ according to Eq.~(\ref{eq:sig_1d}). The dotted line marks the asymptotic value of the $m=2.0$ case as given by Eq.~(\ref{eq:sigma_Zsq_infty}).
      }
      \label{fig:sigZ_1d}
    \end{figure}

    Figure~\ref{fig:sigZ_1d} shows profiles of both quantities for weakly \mbox{($m=1.1$)} and moderately ($m=2$) super-Alfv\'enic flows, qualitatively confirming the shape of functional profiles seen in the simulations (cf.\ Fig.~\ref{fig:sigmaD_runs}).
    At large distances $s \gg 1$,
    \begin{equation}
      \label{eq:sigma_Zsq_infty}
      \lim_{s \rightarrow \infty} \sigma_\mathrm{c}(s) = -\frac{2m}{m^2+1}
      \quad \mbox{and} \quad
      \lim_{s \rightarrow \infty} Z^2(s) = 0
    \end{equation}
    for any $m$ and any $\delta>0$. In particular, this shows that the sign of $\sigma_\mathrm{c}$ is anti-correlated with magnetic polarity (expressible through the sign of $m$, which encodes the orientation of $\vec{V}_\mathrm{\!\!A}$, and hence of $\vec{B}$, with respect to $\vec{U}$) because neither the numerator nor the denominator of Eq.~(\ref{eq:sigma_Zsq_infty}) change sign for $s>1$ and \mbox{$m>1$}.
    
    Furthermore, it can be seen that for low $m$ near unity, $|\sigma_\mathrm{c}|$ grows very quickly even if the flow diverges sub-spherically ($\delta < 2$). This observation may explain the sharp jump in $\sigma_\mathrm{c}$ (from zero on the inner boundary to almost $-1$ in the directly adjacent halo region) visible in the lowermost right panel of Fig.~\ref{fig:ext-vs-ngc}. The near-disk Alfv\'en Mach numbers exhibited by the N4631 run are indeed around unity, and even below unity at distances $R \lesssim 8$\,kpc. This is precisely the radius beyond which a transition region of finite width becomes apparent in said figure.
    
\end{appendix}
\end{document}